\documentclass[a4paper,11pt]{article}
\usepackage{jheppub} 

\title{Solving recurrence relations for multiloop integrals 
in the limit of large values of the dimensional regularization parameter}

\author{P. A. Baikov}
\affiliation{Skobeltsyn Institute of Nuclear Physics, 
Lomonosov Moscow State University,
\\1(2), Leninskie gory, Moscow  119991, Russian Federation}

\emailAdd{baikov@theory.sinp.msu.ru}

\abstract{
A method for calculating the $1/d$ expansion coefficients for solutions 
of integration by parts relations for Feynman integrals is presented.
The idea is to use linear substitutions to transform these relations 
to an explicitly recursive form.
A possible type of such substitutions is proposed for the case of vacuum integrals.
Its applicability is shown for several families of massless (with one massive line) 
vacuum integrals up to the 7-loop level.
}

\arxivnumber{2407.09112}

\begin{document} 
\maketitle
\flushbottom

\section{Introduction}
\label{sec:intro}

Multiloop Feynman integrals evaluation is an important ingredient 
of higher-order perturbative calculations.
A very efficient tool in this sense is integration by parts (IBP)
relations \cite{Chetyrkin:1981qh,Tkachov:1981wb} 
which are used in a large number of specialized computer programs
\cite{Gorishnii:1989gt,Anastasiou:2004vj,Smirnov:2019qkx,Klappert:2020nbg,Lee:2012cn,
Studerus:2009ye,vonManteuffel:2012np,Ruijl:2017cxj}.

As a result, many record-breaking calculations have been performed; 
for a few recent examples, see 
\cite{Georgoudis:2021onj,Chakraborty:2022yan,Lee:2023dtc,Behring:2020cqi,Bern:2023ccb}.
However, extension of the IBP method to higher perturbation orders 
faces significant challenges such as growing computer and hand-working demands, 
increasing the number of subcases that need to be considered and therefore
the probability of errors of various origins.
Although some reserve still exists, both purely mathematical 
and algorithmic improvements are highly desirable for further progress.

The $1/d$ expansion (where $d$ is the dimensional regularization parameter)
in the context of applying the IBP relations was proposed in 
\cite{Baikov:2005nv, Baikov:2007zza} and was
successfully used in the number of calculations (see, for example 
\cite{Baikov:2008jh,Baikov:2009bg,Baikov:2010je,Baikov:2012er,
Baikov:2012zm,Baikov:2012zn,Baikov:2014qja,Baikov:2016tgj,Baikov:2017ujl}).
From a technical point of view this requires expansion 
in the limit $1/d\rightarrow 0$ 
of integrals with integrands depending on $P(x_i)^{d/2}$ where $P(x_i)$ 
is some multivariate polinomial. 
This expansion causes the size of  intermediate expressions to rapidly increase
and calculations to a sufficiently high order of $1/d$ become possible 
only after manual optimization (smart redefinition of variables, etc.), 
which was carried out individually for different subclasses of integrals.
Bearing in mind the significant increase in the 
subclasses number at the next loop(s) level,
in this article we reformulate the problem as constructing solutions to IBP relations
in the form of formal series of $1/d$ and propose linear substitutions 
that transform these relations into an explicitly recursive form.

The plan of the paper is the following. 
In section \ref{sec:FIandIBP} we fix the notations and briefly recall 
the main points of the 
\cite{Baikov:1996rk, Baikov:1996iu,Baikov:2005nv, Baikov:2007zza}. 
Then in section \ref{sec:comb} we introduce (for one-parameter case)
specific linear substitutions, which we apply 
in section \ref{sec:ShiftsA} to the IBP relations.
In the next section \ref{sec:Examp} we consider several examples of 
massless (with one massive line) vacuum integral families up to the 7-loop level.
Finally, in section \ref{sec:Conc} we discuss a possible extension
to more complex cases.

\section{Integration by parts relations for Feynman integrals}
\label{sec:FIandIBP}
Let us consider the $d$-dimensional $L$-loop vacuum Feynman integrals:
\begin{eqnarray}
 F(n_1,\ldots,n_N;d;m^2) =
\int \ldots \int \frac{\mbox{d}^dp_1\ldots \mbox{d}^dp_L}
{D_1^{n_1}\ldots D_N^{n_N}}
\equiv
(m^2)^{Ld/2 - \Sigma n_i} f(n_1,\ldots,n_N;d),
\label{eqbn}
\end{eqnarray}
where $p_i$ (${i=1,\ldots,L}$) are loop momenta and  
\begin{eqnarray}
D_a&=&A^{ij}_a p_i\cdot p_j -\mu_a m^2, \quad a=1,\ldots,N=L(L+1)/2 
\label{props}
\end{eqnarray}
are inverse propagators. The matrix $A^{ij}_a$ is supposed to be symmetrical with respect
to upper indices.
The $1/D_a$ can be raised as to positive powers
(``denominators'' or ``lines'') as non-positive (``numerators'').
Summation over repeating indices is implied and in many cases
(unless it's confusing) it will be omitted.
In what follows we will set $m^2=1$; 
we will also omit $d$ in the arguments and use the shorthand $(n_1,\ldots,n_N)\equiv (\underline{n})$.

We assume that the set~(\ref{props}) is complete and we can express
scalar products $p_i\cdot p_j$ through $D_a$:
\begin{eqnarray}
p_i\cdot p_k &=& \hat{A}^{ik}_a(D_a+\mu_a)\,. 
\label{pp2D}
\end{eqnarray}
The matrix $\hat{A}^{ij}_a$ is also symmetrical with respect
to $(ij)$ and is inverse to $A^{kl}_a$ in the sense that 
\begin{eqnarray}
\hat{A}^{ij}_a A^{kl}_a&=&(\delta_{ik} \delta_{jl}+\delta_{il} \delta_{jk})/2,\quad
A^{ik}_a \hat{A}^{ik}_b =\delta_{ab}.
\label{AAdef}
\end{eqnarray}
Please note that in \cite{Baikov:2005nv} for the coefficients in (\ref{pp2D})
we used the notation $\tilde{A}$; we change it to $\hat{A}$ because we want to reserve 
the ``$\sim$'' sign for ``soft'' objects (see section \ref{sec:ShiftsA}).

Let's define
\begin{eqnarray}
{\bf I}^-_a f(\ldots, n_a,\ldots )&\equiv& f(\ldots, n_a-1,\ldots ),
\nonumber \\
{\bf I}^+_a f(\ldots, n_a,\ldots )&\equiv& n_a f(\ldots, n_a+1,\ldots)
\nonumber
\end{eqnarray}
(no summation in $n_a$ in the last relation).

IBP relations can be obtained \cite{Chetyrkin:1981qh,Tkachov:1981wb}
by inserting the operator 
$(\partial/\partial p_i)\cdot p_k$ 
in the integrand, 
equating the integral of the derivative to zero
and expressing all the terms resulting from the differentiation through the 
integrals with shifted $(\underline{n})$. 
Starting from eq.~(\ref{eqbn}) we obtain (up to $1/2$ factor)
\begin{eqnarray}
0=R_{ik}(\underline{n})&=&\big( (d/2)\,\delta_{ik} 
- \hat{A}^{kl}_a ({\bf I}^-_a+\mu_a)
A_b^{il}{\bf I}_b^+\big) \, 
f(\underline{n}) \,,
\label{rr0}
\\
 &=&\big( (d-L-1)/2 \,\delta_{ik}
- A_b^{il}{\bf I}_b^+ \, 
\hat{A}^{kl}_a ({\bf I}^-_a+\mu_a) \big)
f(\underline{n}) \,,
\label{rr1}
\end{eqnarray}
where we use
$[{\bf I}^-_a,{\bf I}^+_b]=\delta_{ab}$ to derive eq.~(\ref{rr1}).

A linear combination of $R_{ik}(\underline{n})$ with shifted
$(\underline{n})$ corresponds to shifts of $(\underline{n})$ in $f(\underline{n})$
and hence to the insertion of some combination of 
$({\bf I}^-_a, {\bf I}^+_a)$ just before $f(\underline{n})$.

In many practically important cases eq.~(\ref{rr0}) can be used to relate 
integrals with arbitrary values of $(\underline{n})$ to a linear combination
of a finite number of irreducible (master) integrals: 
\begin{eqnarray}
f(\underline{n})&=&\sum_k c_k(\underline{n})\,M_k,\quad
M_k=f(\underline{n}_k),
\label{f2M}
\end{eqnarray}
where $f(\underline{n}_k)$ are chosen as the most convenient for 
direct calculation, usually with values $(\underline{n}_k)$ 
from the set $(-1,0,+1)$.

Such a reduction can be achieved as by deriving relations with an explicitly 
recursive structure (MINCER \cite{Gorishnii:1989gt} for 3-loop propagator integrals,
Forcer \cite{Ruijl:2017cxj} for the next loop level) 
as by Laporta approach \cite{Laporta:2000dsw} with a number of implementations
\cite{Anastasiou:2004vj,Smirnov:2019qkx,Klappert:2020nbg,Studerus:2009ye,vonManteuffel:2012np}. 
In any case, reduction to irreducible integrals
during actual calculations requires huge computer resources, so any progress in this field
is very desirable.
In this context, a direct calculation of c(\underline{n}) was proposed
\cite{Baikov:1996rk, Baikov:1996iu}.
Indeed, if $M_k$ are linearly independent, then 
$c_k(\underline{n})$ are rational in $d$ solutions of IBP relations
and can be obtained from any other solutions set as linear combinations  
satisfying the condition $c_i(\underline{n}_k)=\delta_{ik}$.

There are natural boundary conditions on $c_k(\underline{n})$,
which follow from the assumption that no additional denominators (lines) appear
on the right side of eq.~(\ref{f2M})
(since the purpose of the reduction procedure is to relate the original integral
with simpler ones):
\begin{eqnarray}
c_k(\underline{n})&\neq& 0\quad \mbox{only if}\,\,n_a>0\,\,\mbox{for all}\,\, a\in H_k,
\nonumber
\end{eqnarray}
where $ H_k=\{a_1,\dots,a_{N_k}\}$ is the set of line numbers of the irreducible integral $M_k$
and $N_k$ is the total number of lines in $M_k$.
In some cases we will also use an alternative boundary condition for $a\notin H_k$:
\begin{eqnarray}
f(\underline{n})&\neq& 0\quad \mbox{only if}\,\,n_a<1\,\,\mbox{for all}\,\, a\notin H_k.
\label{bcondnum}
\end{eqnarray}
Here and below we consider $f(\underline{n})$ as some solution of 
recurrence relations (and integrals (\ref{eqbn}) as a particular example).

In \cite{Baikov:1996rk, Baikov:1996iu} $c_k(\underline{n})$ were constructed using 
the auxiliary integral representation
\begin{eqnarray}
f(\underline{n})&=&\int dx_1 \dots dx_N\, f(\underline{x})/(x_1^{n_1}..\ x_N^{n_N}),
\label{intrepr}
\end{eqnarray}
where $f(\underline{x})$ should be chosen according to recurrence relations;
for example, relations for integrals (\ref{eqbn}) require
\begin{eqnarray}
f(\underline{x})=\det(\hat{A}^{ik}_a (x_a+\mu_a))^{(d-L-1)/2}.
\label{fvac}
\end{eqnarray}
Integrals (\ref{intrepr}) can be expanded in the $1/d \rightarrow 0$ limit
\cite{Baikov:2005nv, Baikov:2007zza}
but as mentioned in the introduction this is difficult to extend to the next loop(s) level.
Therefore, as a more promising alternative, we are going 
to explore ways to find the solutions to IBP relations
as formal series in $1/d$.

\section{Linear substitutions in recurrence relations: one-parameter case}
\label{sec:comb}
Let's consider the one-parameter recurrence relation:
\begin{eqnarray}
0&=&R(n)=R({\bf I}_f^-,{\bf I}_f^+) f(n),
\label{RRapp}
\end{eqnarray}
where by definition
\begin{eqnarray}
{\bf I}_f^+ f(n) &\equiv& n f(n+1),\quad 
{\bf I}_f^- f(n) \equiv f(n-1),
\label{If}
\end{eqnarray}
and hence
\begin{eqnarray}
\big[{\bf I}_f^-, {\bf I}_f^+\big] &\equiv&
{\bf I}_f^- {\bf I}_f^+ - {\bf I}_f^+ {\bf I}_f^- = 1,
\nonumber \\
\big[{\bf I}_f^-, ({\bf I}_f^+)^k\big] & = & k\,({\bf I}_f^+)^{k-1}, \quad
\big[{\bf I}_f^+, ({\bf I}_f^-)^k\big] = - k\,({\bf I}_f^-)^{k-1},
\nonumber \\
\big[{\bf I}_f^-, S({\bf I}_f^+)\big] & = & 
\partial_{{\bf I}_f^+}S({\bf I}_f^+), \quad
\big[{\bf I}_f^+, S({\bf I}_f^-)\big] = 
-\partial_{{\bf I}_f^-}S({\bf I}_f^-),
\nonumber
\end{eqnarray}
where $S({\bf I})$ is any formal series in ${\bf I}$.

We are going to study the transformation of eq.~(\ref{RRapp})
after linear redefinition of $f(n)$.
The idea is to replace $f(n)$ with some combination of $g(n)$ to obtain
a simpler recurrence relation for $g(n)$, 
solve it and re-express the results in terms of $f(n)$.

Let's start with eq.~(\ref{RRapp}) with the boundary condition
\begin{eqnarray}
f(n<1)=0.
\label{cond1}
\end{eqnarray}
Due to eq.~(\ref{cond1}) if $k>n-1$ then $({\bf I}_f^-)^kf(n)=f(n-k)=0$, so
\begin{eqnarray}
g(n)&=&
e^{\lambda {\bf I}_f^-} f(n)
\equiv 
\sum_{k=0}^{\infty} \frac{(\lambda {\bf I}_f^-)^k}{k!}f(n)
\label{gtof}
\end{eqnarray}
will be equal to the finite linear combination of $f(k)$ with $k=1,\dots,n$.
We are going to derive the recurrence relation for $g(n)$, which follows from eq.~(\ref{RRapp}).
To do this we need to find the relations between
operators ${\bf I}_f^\pm$ defined by eq.~(\ref{If})
and ${\bf I}_g^\pm$, acting in the same way on $g(n)$.
Let's start with ${\bf I}_g^+$:
\begin{eqnarray}
{\bf I}_g^+ g(n)&\equiv&n g(n+1)
= n e^{\lambda {\bf I}_f^-} f(n+1)
= e^{\lambda {\bf I}_f^-} n f(n+1) 
= e^{\lambda {\bf I}_f^-} {\bf I}_f^+ f(n) 
\nonumber\\
&=& \big( \big[ e^{\lambda {\bf I}_f^-}, {\bf I}_f^+ \big]
+ {\bf I}_f^+ e^{\lambda {\bf I}_f^-} \big) f(n) 
= \big( \partial_{{\bf I}_f^-} (e^{\lambda {\bf I}_f^-})
+ {\bf I}_f^+ e^{\lambda {\bf I}_f^-}\big) f(n) 
= (\lambda+ {\bf I}_f^+ ) e^{\lambda {\bf I}_f^-} f(n) 
\nonumber\\
&=& ({\bf I}_f^+ + \lambda ) g(n).
\nonumber
\end{eqnarray}
The relation for ${\bf I}_g^-$ is more simple:
\begin{eqnarray}
{\bf I}_g^- g(n)& \equiv & g(n-1)
= e^{\lambda {\bf I}_f^-} f(n-1)
= e^{\lambda {\bf I}_f^-} {\bf I}_f^- f(n) 
= {\bf I}_f^- e^{\lambda {\bf I}_f^-} f(n) 
\nonumber\\
&=& {\bf I}_f^-  g(n).
\nonumber
\end{eqnarray}
Now we can invert eq.~(\ref{gtof}):
\begin{eqnarray}
f(n)&=&e^{- \lambda {\bf I}_f^-} g(n)
\nonumber\\
    &=&e^{- \lambda {\bf I}_g^-} g(n). 
\label{ftog}
\end{eqnarray}
Again, due to eqs.~(\ref{cond1},~\ref{gtof})
it turns out that $g(n<1)=0$ and eq.~(\ref{ftog}) consists of a finite number of terms.

The recurrence relation for $g(n)$ follows from the linear combination 
of eqs.~(\ref{RRapp}) with shifted $n$ (${\bf I}_R^- R(n)\equiv R(n-1)$):
\begin{eqnarray}
0&=&e^{\lambda {\bf I}_R^-} R(n)
= R({\bf I}_f^-, {\bf I}_f^+) e^{\lambda {\bf I}_f^-} f(n)
= R({\bf I}_f^-, {\bf I}_f^+) g(n) 
\nonumber \\
&=&R({\bf I}_g^-,{\bf I}_g^+ - \lambda) g(n).
\nonumber
\end{eqnarray}

The case with the boundary condition 
\begin{eqnarray}
f(n>0)=0
\label{cond2}
\end{eqnarray}
is considered similarly, we present the final formulas:
\begin{eqnarray}
g(n) & = & e^{\lambda {\bf I}_f^+} f(n), \quad g(n>0)=0,  
\nonumber \\
f(n) & = & e^{-\lambda {\bf I}_g^+} g(n), 
\label{ftog0} 
\\
{\bf I}_g^-&=&{\bf I}_f^- -\lambda, \quad {\bf I}_g^+={\bf I}_f^+, 
\nonumber \\
0 & = & R({\bf I}_g^- +\lambda, {\bf I}_g^+) g(n).
\label{RRg}
\end{eqnarray}

The boundary condition (\ref{cond2}) corresponds to the condition
(\ref{bcondnum}), 
that is, the case when the master integral $M_k$ does not have the line $D_a$.
However, there may be such a line on the left side of eq.~(\ref{f2M}), 
which means that $n_a$ in $c_k(\underline{n})$ can be positive.
So, let us extend the substitution (\ref{ftog0}) to positive $n$.
For $n<1$ we can rewrite $g(n) = ({\bf I}_g^-)^{-n} g(0)$
and then eq.~(\ref{ftog0}) as follows:
\begin{eqnarray}
f(n) & = & e^{-\lambda {\bf I}_g^+} ({\bf I}_g^-)^{-n} g(0)
 = ({\bf I}_g^- +\lambda ) e^{-\lambda {\bf I}_g^+} ({\bf I}_g^-)^{-n-1}  g(0)
 =\dots
= ({\bf I}_g^- +\lambda )^{-n} e^{-\lambda {\bf I}_g^+} g(0)
\nonumber \\
 & = & ({\bf I}_g^- +\lambda )^{-n} g(0).
\label{ftog1} 
\end{eqnarray}

For $f(n<1)$ the binomial expansion reproduces the finite sum (\ref{ftog0}) as it should.
For $f(n>0)$ we expand the right side of eq.~(\ref{ftog1})
into a formal series in ${\bf I}_g^-$. From the negative binomial series we get
\begin{eqnarray}
f(n>0) & = & ({\bf I}_g^- +\lambda )^{-n} g(0) 
\nonumber \\
       & = & \sum_{k=0}^{\infty} (-1)^k \frac{(n+k-1)!}{(n-1)!\,k!}
 \lambda^{-n-k} ({\bf I}_g^-)^{k} g(0)
\nonumber \\
       & = & \sum_{k=0}^{\infty} (-1)^k \frac{(n+k-1)!}{(n-1)!\,k!} \lambda^{-n-k} g(-k),
\label{ftog2}
\end{eqnarray}
that is the infinite sum of $g(n<1)$.
However, as we will see in section \ref{sec:ShiftsA}
one can choose $\lambda$ so that the $k$-term in the sum (\ref{ftog2})
has leading asymptotics $d^{-k/2}$
(see eq.~(\ref{ftogFI})),
therefore, by keeping enough terms in this sum 
and assuming that $g(n<1)$ satisfies eq.~(\ref{RRg}) 
we can satisfy (\ref{RRapp}) in any desired order in $1/d $.
As a result, $f(n)$ can be parameterized using $g(k<1)$
for any value of $n$.\footnote{Note that a direct extension of eq.~(\ref{ftog0}) for $n>0$
produces an infinite sum over $k$ of terms with undesirable leading asymptotic $d^{k/2}$ (see eq.~(\ref{ftogFI})).}
In what follows, by the expression (\ref{ftog0}) in the case (\ref{cond2}) 
we will mean the extension (\ref{ftog2}) for $n>0$
(in the $\lambda=0$ case when (\ref{ftog2}) is not defined we will set $f(n)=g(n)$).

Finally, if we rescale $f(n)=\alpha^{n}g(n)$, then eq.~(\ref{RRapp})
transforms into
\begin{eqnarray}
0&=&R({\alpha^{-1} \bf I}_g^-,\alpha {\bf I}_g^+) g(n).
\nonumber
\end{eqnarray}
  
\section{Linear substitutions in recurrence relations: Feynman integrals}
\label{sec:ShiftsA}

Let us now return to the general case
and construct, if possible, a solution to eq.~(\ref{rr0}) for some
set of propagator indices $H_k=\{a_1,\dots ,a_{N_k}\}$.
We will call the index $a \in H_k$ ``hard''.
(substitutions (\ref{ftog}) will be applied), and
$a \notin H_k$ ``soft' (substitutions (\ref{ftog0},~\ref{ftog2})).

For any indexed object $x_a$ we define projections onto the ``hard'' and ``soft'' subspaces:
${\overline{x}}_a \equiv x_{\overline{a}}=$ (if $a\in H_k$ then $x_a$ else $0$) 
and
${\tilde{x}}_a \equiv x_{\tilde{a}}= x_a - {\overline{x}}_a$;
as the result $x_a=\overline{x}_a + \tilde{x}_a$.
For $x_a={\bf I}_a^{\pm}$ we have
${\bf I}_a^{\pm}=
{\tilde{\bf I}}_a^{\pm}+
{\overline{\bf I}}_a^{\pm},\,
\big[ {\tilde{\bf I}}_a^{\pm}, {\overline{\bf I}}_b^{\pm} \big] = 0.
$

Let us redefine 
\begin{eqnarray}
d/2&=&1/\alpha^2, \quad \alpha\rightarrow 0.
\nonumber
\end{eqnarray}
According to section \ref{sec:comb} the substitutions
\begin{eqnarray}
f(\underline{n})&=&
\exp(\alpha^{-2}\overline{\lambda}_a\, \overline{{\bf I}}_{f^{\prime},a}^- -
\tilde{\lambda}_a \tilde{\bf I}_{f^{\prime},a}^+) 
f^{\prime}(\underline{n}), 
\nonumber \\
f^{\prime}(\underline{n})&=&\alpha^{N_k-\sum_a n_a}
g(\underline{n})
\label{ftogFI}
\end{eqnarray}
will change
\begin{equation}
\overline{{\bf I}}_{f,a}^+ =
\alpha^{-2}(\alpha\overline{{\bf I}}_{g,a}^+ +\overline{\lambda}_a),\quad
\tilde{{\bf I}}_{f,a}^- =
\alpha \tilde{{\bf I}}_{g,a}^- +\tilde{\lambda}_a,\quad
\tilde{I}_f^+=\alpha^{-1}\tilde{I}_g^+,\quad
\overline{I}_f^-=\alpha \overline{I}_g^-,
\nonumber
\end{equation}
and therefore eq.~(\ref{rr0}) (multiplied by $\alpha^2$) will be transformed into
\begin{eqnarray}
0 &=& 
\Big(\delta_{ik} 
- \hat{A}^{kl}(\alpha\,{\bf I}^-_{g} + \tilde{\lambda} + \mu)
A^{li}\big({\alpha\,\bf I}_{g}^+ +\overline{\lambda} \big)
\Big) \, g(\underline{n}) \,
\nonumber\\
&=& \Big( \delta_{ik} 
- \hat{A}^{kl}(\tilde{\lambda} + \mu) A^{li}\,\overline{\lambda} 
\nonumber\\
&\phantom{=}& 
- \alpha\,
\big(
\hat{A}^{kl}(\tilde{\lambda} + \mu)
A^{li}{\bf I}_{g}^+
+\hat{A}^{kl}{\bf I}^-_{g}
A^{li}\overline{\lambda}
\big)
- \alpha^2 \hat{A}^{kl}{\bf I}^-_{g}
A^{li}{\bf I}_{g}^+
\Big) g(\underline{n}).
\label{rr1g}
\end{eqnarray}
Here and in some places below we omit the ``propagator'' indices 
over which the summation is performed, assuming that matrices $A, \hat{A}$ 
without a subscript are summed with the object located on the right.
Also, compared to eq.~(\ref{rr0}), we have changed the order of the 
upper indices of the matrix $A^{kl}_a$ (since it is symmetrical).

Eqs.~(\ref{rr1g}) generate relations for the expansion coefficients $g_j(\underline{n})$ 
in \hbox{$g(\underline{n}) =\sum_{j=0}^{\infty} \alpha^j\ g_j(\underline{n})$}:
\begin{eqnarray}
0&=& \big( \delta_{ik} 
- \hat{A}^{kl}(\tilde{\lambda} + \mu) A^{li}\overline{\lambda} 
\big) g_j(\underline{n})
\nonumber\\
&\phantom{=}& 
- \big(
\hat{A}^{kl}(\tilde{\lambda} + \mu)
A^{li}{\bf I}^+
+\hat{A}^{kl}{\bf I}^-
A^{li}\overline{\lambda}
\big) g_{j-1}(\underline{n})
\nonumber\\
&\phantom{=}& 
- \hat{A}^{kl}{\bf I}^-
A^{li}{\bf I}^+
g_{j-2}(\underline{n}).
\label{rr1ge}
\end{eqnarray}
From here on we omit the indices $f, f^{\prime}, g$ in ${\bf I}^{\pm}$ 
if the ``default'' action is implied.

Equations for $g_0(\underline{n})$
do not contain ${\bf I}_a^\pm$ operators and in general lead to $g_0(\underline{n})=0$
(and as a result $g_j(\underline{n})=0$),
therefore, to obtain non-trivial solutions it is necessary to choose
$\lambda_a$ satisfies the equations
\begin{eqnarray}
\delta_{ik}&=& \hat{A}^{kl}(\tilde{\lambda} + \mu) A^{li}\,\overline{\lambda}
\label{condla}.
\end{eqnarray}

Eqs.~(\ref{rr1ge}) with the condition (\ref{condla}) become simpler:
\begin{eqnarray}
0&=&
\big(
\hat{A}^{kl}(\tilde{\lambda}+\mu)A^{li}{\bf I}^+
+\hat{A}^{kl}{\bf I}^-A^{li}\overline{\lambda}
\big)g_{j}(\underline{n})
\nonumber\\
&\phantom{=}& 
+\hat{A}^{kl}{\bf I}^-A^{li}{\bf I}^+
g_{j-1}(\underline{n}).
\label{rr1ge1}
\end{eqnarray}

We are going to use eqs.~(\ref{rr1ge1}) to reduce
$g_{j}(\underline{n})$ (up to $g_{j-1}(\underline{n})$ terms)
to ``corner'' values: ``hard'' $\overline{n}_a \to 1$
and ``soft'' $\tilde{n}_a \to 0$
(assuming ``soft'' $\tilde{\lambda}_a \neq 0$, see the end of section~\ref{sec:comb}).
This would be possible if 
the $g_{j}(\underline{n})$ part of eqs.~(\ref{rr1ge1}) could be solved explicitly 
with respect to the operators ($\overline{\bf I}_a^+, \tilde {\bf I}_a^-$), 
increasing the distance from the corner.
Let us multiply eqs.~(\ref{rr1ge1}) by
$(\hat{A}_{a} A\overline{\lambda})^{ik}$:
\begin{eqnarray}
0&=&
\Big( 
\mbox{Tr} 
\big( 
\hat{A}_{a}A\overline{\lambda}\,\hat{A}(\tilde{\lambda} + \mu)A{\bf I}^+
\big)
+\mbox{Tr}
\big( 
\hat{A}_{a}A\overline{\lambda}\,\hat{A}{\bf I}^- A\overline{\lambda}
\big)
\Big) g_{j}(\underline{n})
+ \mbox{Tr}
( 
\dots
) g_{j-1}(\underline{n})
\nonumber\\
&=& 
\big( 
{\bf I}_a^+
+\mbox{Tr}
( 
\hat{A}_{a}A\overline{\lambda}\,\hat{A}_b A\overline{\lambda}
)
{\bf I}_b^-
\big) g_{j}(\underline{n})
+ \mbox{Tr}
( 
\hat{A}_{a}A\overline{\lambda}\,\hat{A}{\bf I}^-A{\bf I}^+) 
g_{j-1}(\underline{n}).
\label{rr1ge2}
\end{eqnarray}
Here we omit the repeated ``loop'' indices, $\mbox{Tr}(\cdot)$ means the contraction of outer indices $(ik)$;
we also use the condition~(\ref{condla}) and $\mbox{Tr}(\hat{A}_a A_b)=\delta_{ab}$.
In the case of ``soft'' $\tilde{a}$, the $g_{j}(\underline{n})$ part of eqs.~(\ref{rr1ge2}) depends on
increasing $\tilde{\bf I}_a^-$ and decreasing $(\tilde{\bf I}_a^+,\,\overline{\bf I}_a^-)$
operators and if
\begin{eqnarray}
\det_{\tilde{a}\tilde{b}} \mbox{Tr}(\hat{A}_{\tilde{a}} A\overline{\lambda}\,
\hat{A}_{\tilde{b}} A\overline{\lambda})
&\neq& 0,
\label{condlass1}
\end{eqnarray}
where both indices $a, b$ are ``soft'', we can
express increasing ``soft'' operators $\tilde{\bf I}_a^-$ in terms of decreasing ones:
\begin{eqnarray}
\tilde{\bf I}_a^- &\rightarrow& (\tilde{\bf I}_a^+,\,\overline{\bf I}_a^-). 
\label{soft2decr}
\end{eqnarray}

For ``hard'' $\overline{a}$ eqs.~(\ref{rr1ge2}) explicitly express the remaining increasing operators
$\overline{\bf I}_a^+$ in terms of $(\tilde{\bf I}_a^-,\,\overline{\bf I}_a^-)$
and due to eqs.~(\ref{soft2decr}) in terms of decreasing
$(\tilde{\bf I}_a^+,\,\overline{\bf I}_a^-)$ only, so as the result we have an explicit
reduction to ``corner'' values
\begin{eqnarray}
(\tilde{\bf I}_a^-,\,\overline{\bf I}_a^+)
&\rightarrow&
(\tilde{\bf I}_a^+,\,\overline{\bf I}_a^-).
\label{incr2decr}
\end{eqnarray}

The $L(L+1)/2$ equations~(\ref{incr2decr}) 
are linear combinations (by means of eqs.~(\ref{rr1ge2})) of the $L^2$ eqs.~(\ref{rr1ge1}),
so we can expect additional linear
relations on ${\bf I}_a^\pm$, which can lead to a contradiction.
Fortunately if (\ref{condlass1}) and hence (\ref{incr2decr}) are valid
this is not the case. 
Indeed, eqs.~(\ref{rr1ge2}) are the combinations of eqs.~(\ref{incr2decr}).
Then the $g_{j}(\underline{n})$ parts of eqs.~(\ref{rr1ge1})
are the linear combinations of the $g_{j}(\underline{n})$ parts of eqs.~(\ref{rr1ge2})
with the coefficients $(\hat{A}(\tilde{\lambda}+\mu)A_{a})^{ki}$, 
so substituting (\ref{incr2decr}) into eqs.~(\ref{rr1ge1})
nullifies this part and no additional relations appear.

The case when some of the “soft” $\tilde{\lambda}_a = 0$ 
is more complicated, since the corresponding $\tilde{n}_a$ can be both negative 
and positive and finding out the possibility of reducing them to a minimum value
requires a more detailed study of the $g_{j}(\underline{n})$ parts of eqs.~(\ref{rr1ge1}).

Eqs.~(\ref{condla}) lead to several useful consequences.
First, $\mbox{Tr}(\cdot )$ of eqs.~(\ref{condla}) due to eq.~(\ref{AAdef}) leads to
\begin{eqnarray}
L&=& \overline{\mu}_a\,\overline{\lambda}_a.
\label{Trcondla}
\end{eqnarray}

Then, $\det(\cdot )$ of eqs.~(\ref{condla}) leads to 
\begin{eqnarray}
1&=&\det \big(\hat{A}(\tilde{\lambda}+\mu)\big)\det (A\overline{\lambda})
\nonumber \\
&\Rightarrow & \det \big(\hat{A}(\tilde{\lambda}+\mu)\big)\neq 0.
\label{detnot0}
\end{eqnarray}

Next, $\hat{A}(\tilde{\lambda}+\mu )$ and
$A\overline{\lambda}$
considered as matrices with indices $(ik)$ are reciprocal
\begin{eqnarray}
A\overline{\lambda}&=&\hat{A}(\tilde{\lambda}+\mu )^{-1}.
\label{A2hatA}
\end{eqnarray}

Finally, let us multiply eqs.~(\ref{condla}) by
$\big(\hat{A}_a \hat{A}(\tilde{\lambda}+\mu)^{\mbox{\small Adj}}\big)^{ik}$:
\begin{eqnarray}
\partial_a \det \big(\hat{A}(\tilde{\lambda}+\mu)\big)
&=& \overline{\lambda}_a \,
\det \big(\hat{A}(\tilde{\lambda}+\mu)\big),
\label{condlas}
\end{eqnarray}
where $A^{\mbox{\small Adj}}\equiv\det (A)\,A^{-1}$,
$\partial_a \det (A \lambda)\equiv \partial_{x_a} \det (A^{ik}_a x_a) |_{x_a=\lambda_a}$.
For ``soft'' $\tilde{a}$ eqs.~(\ref{condlas}) give
\begin{eqnarray}
 \partial_{\tilde{a}} \det (\hat{A}(\tilde{\lambda}+\mu))&=&0.
\label{condlas1}
\end{eqnarray}

Using eqs.~(\ref{A2hatA}) we can re-express eq.~(\ref{condlass1}) via $\tilde{\lambda}_a$:
\begin{eqnarray}
0&\neq&\det_{\tilde{a}\tilde{b}} \mbox{Tr}\big(\hat{A}_{\tilde{a}} A\overline{\lambda}\,
\hat{A}_{\tilde{b}} A\overline{\lambda}\big)
\nonumber\\
&=&
\det_{\tilde{a}\tilde{b}} \Big(\mbox{Tr}\big( 
\hat{A}_{\tilde{a}}
\hat{A}(\tilde{\lambda}+\mu )^{-1}
\hat{A}_{\tilde{b}}
\hat{A}(\tilde{\lambda}+\mu )^{-1}
\big)\Big)
\nonumber\\
&\propto&
\det_{\tilde{a}\tilde{b}} \Big(\partial_{\tilde{a}} \partial_{\tilde{b}} 
\det \big(\hat{A}(\tilde{\lambda}+\mu)\big) /\det \big(\hat{A}(\tilde{\lambda}+\mu)\big)\Big),
\label{condlass2}
\end{eqnarray}
where in the third line we use eqs.~(\ref{condlas1})
and standard matrix formulas
\begin{eqnarray}
\partial_x \det(A)&=&\det(A)\,\mbox{Tr}\big(A^{-1}\partial_x A\big),\quad
\partial_x(A^{-1}) = -A^{-1}\,\partial_x(A)\,A^{-1}.
\nonumber
\end{eqnarray}

Please note that eqs.~(\ref{condlas1}, \ref{detnot0}, \ref{condlass2})
correspond to eqs.~(16) from \cite{Baikov:2005nv}, that is to the conditions 
for the existence of the $1/d$ expansion of the integral representation 
(\ref{intrepr}, \ref{fvac}).

\section{Examples}
\label{sec:Examp}

Let us consider 2-loop example with inverse propagators
\begin{equation}
D_1=p_1^2,\quad D_2=p_2^2,\quad D_3=(p_1+p_2)^2-1.
\nonumber
\end{equation}
Recurrence relations (\ref{rr0}) are
\begin{eqnarray}
R_{11}(\underline{n})&=&1/2 
(d - 2 \overline{\bf I}_1^- \overline{\bf I}_1^+  - \overline{\bf I}_1^- \overline{\bf I}_3^+  + \overline{\bf I}_2^- \overline{\bf I}_3^+  - \overline{\bf I}_3^- \overline{\bf I}_3^+  - \overline{\bf I}_3^+ )
f(\underline{n}),\nonumber\\
R_{12}(\underline{n})&=&1/2 
(\overline{\bf I}_1^- \overline{\bf I}_1^+  + \overline{\bf I}_1^- \overline{\bf I}_3^+  + \overline{\bf I}_2^- \overline{\bf I}_1^+  - \overline{\bf I}_2^- \overline{\bf I}_3^+  - \overline{\bf I}_3^- \overline{\bf I}_1^+  - \overline{\bf I}_3^- \overline{\bf I}_3^+  - \overline{\bf I}_1^+  - \overline{\bf I}_3^+ )
f(\underline{n}),\nonumber\\
R_{21}(\underline{n})&=&1/2 
(\overline{\bf I}_1^- \overline{\bf I}_2^+  - \overline{\bf I}_1^- \overline{\bf I}_3^+  + \overline{\bf I}_2^- \overline{\bf I}_2^+  + \overline{\bf I}_2^- \overline{\bf I}_3^+  - \overline{\bf I}_3^- \overline{\bf I}_2^+  - \overline{\bf I}_3^- \overline{\bf I}_3^+  - \overline{\bf I}_2^+  - \overline{\bf I}_3^+ )
f(\underline{n}),\nonumber\\
R_{22}(\underline{n})&=&1/2 
(d + \overline{\bf I}_1^- \overline{\bf I}_3^+  - 2 \overline{\bf I}_2^- \overline{\bf I}_2^+  - \overline{\bf I}_2^- \overline{\bf I}_3^+  - \overline{\bf I}_3^- \overline{\bf I}_3^+  - \overline{\bf I}_3^+ )
f(\underline{n}).
\label{l2rr}
\end{eqnarray}
Let us assume that all indices are ``hard'',
\begin{equation}
f(\underline{n})=0\,\,\, \mbox{if any of}\,\, (n_1, n_2, n_3) < 1,
\label{f2lcond}
\end{equation}
and substitute (remember that $d/2=\alpha^{-2}$):
\begin{eqnarray}
f(\underline{n})&=&\exp(\alpha^{-2}\sum_{a=1}^3 \overline{\lambda}_a \overline{\bf I}_a^-)
f^{\prime}(\underline{n}),\quad f^{\prime}(\underline{n})= \alpha^{3-\sum_{a=1}^3 n_a} g(\underline{n}), 
\label{ftog2l} \\
f^{\prime}(\underline{n}),\,g(\underline{n})&=&0\,\,\, \mbox{if any of}\,\, (n_1, n_2, n_3) < 1.
\label{g2lcond}
\end{eqnarray}
These substitutions correspond to
\begin{equation}
\overline{{\bf I}}_{f,a}^+ =
\alpha^{-1}\overline{{\bf I}}_{g,a}^+ +\alpha^{-2}\overline{\lambda}_a,\quad
\overline{{\bf I}}_{f,a}^- =\alpha\overline{{\bf I}}_{g,a}^-.
\nonumber
\end{equation}
The $\alpha^{-2}$ parts of (transformed) eqs.~(\ref{l2rr}) lead to equations on $\overline{\lambda}_a$ 
with the solution 
\begin{equation}
\overline{\lambda}_1=-2,\quad \overline{\lambda}_2=-2,\quad \overline{\lambda}_3=2, 
\label{lav}
\end{equation}
while the $\alpha^{-1}$ and $\alpha^0$ parts of the equations $\{0=R_{11}, 0=R_{12}, 0=R_{21}\}$ 
(after substituting eqs.~(\ref{lav}) and multiplying by $\alpha$) lead to recursive relations
\begin{eqnarray}
\overline{\bf I}_1^+  g(\underline{n})&=&\big( - 2 \overline{\bf I}_1^- - 6 \overline{\bf I}_2^- + 2 \overline{\bf I}_3^- + \alpha\,(3 \overline{\bf I}_1^- \overline{\bf I}_1^+  + 2 \overline{\bf I}_1^- \overline{\bf I}_3^+  + \overline{\bf I}_2^- \overline{\bf I}_1^+  - 2 \overline{\bf I}_2^- \overline{\bf I}_3^+  - \overline{\bf I}_3^- \overline{\bf I}_1^+ )\big)
g(\underline{n}),\nonumber\\
\overline{\bf I}_2^+  g(\underline{n})&=&\big( - 6 \overline{\bf I}_1^- - 2 \overline{\bf I}_2^- + 2 \overline{\bf I}_3^- + \alpha\,(2 \overline{\bf I}_1^- \overline{\bf I}_1^+  + \overline{\bf I}_1^- \overline{\bf I}_2^+  +\overline{\bf I}_2^- \overline{\bf I}_2^+  - \overline{\bf I}_3^- \overline{\bf I}_2^+ )\big)
g(\underline{n}),\nonumber\\
\overline{\bf I}_3^+  g(\underline{n})&=&\big(2 \overline{\bf I}_1^- + 2 \overline{\bf I}_2^- - 2 \overline{\bf I}_3^- + \alpha\,( - 2 \overline{\bf I}_1^- \overline{\bf I}_1^+  - \overline{\bf I}_1^- \overline{\bf I}_3^+  +\overline{\bf I}_2^- \overline{\bf I}_3^+  - \overline{\bf I}_3^- \overline{\bf I}_3^+ )\big)
g(\underline{n}).
\label{l2rr1}
\end{eqnarray}

Eqs.~(\ref{ftog2l}-\ref{l2rr1}) allow us to derive a relation between $f(\underline{n})$ 
for a particular $(\underline{n})$ and $f(1,1,1)$; as an example, let us do this for $f(2,1,1)$.
In the first step we need to express $f(2,1,1)$ and $f(1,1,1)$ in terms of $g(\underline{n})$.
According to eq.~(\ref{ftog2l}) with~(\ref{lav}) 
\begin{eqnarray}
f(2,1,1)&=&(1 - 2\alpha^{-2} \overline{\bf I}_1^- +\dots) f^{\prime}(2,1,1) \nonumber\\ 
&=&f^{\prime}(2,1,1) -2\alpha^{-2}f^{\prime}(1,1,1) \nonumber\\ 
&=&\alpha^{-1} g(2,1,1) -2\alpha^{-2} g(1,1,1),
\label{ftog2l2}\\
f(1,1,1)&=&g(1,1,1). 
\label{ftog2l1}
\end{eqnarray}
From here on, dots denote terms that do not contribute due to the conditions (\ref{f2lcond}, \ref{g2lcond}).
In the second step we reduce the right-hand side of eq.~(\ref{ftog2l2}) to $g(1,1,1)$. 
The first equation (\ref{l2rr1}) for $(\underline{n})=(1,1,1)$ gives
\begin{equation}
g(2,1,1)=(3\alpha  \overline{\bf I}_1^- \overline{\bf I}_1^+  + \dots ) g(1,1,1)=3\alpha g(1,1,1),
\label{gtog2l}
\end{equation}
and hence
\begin{equation}
f(2,1,1)=(3-2\alpha^{-2}) g(1,1,1)=(3-d) g(1,1,1). 
\label{ftog2l21}
\end{equation}
From eqs.~(\ref{ftog2l1}, \ref{ftog2l21}) we finally obtain
\begin{equation}
f(2,1,1)=(3-d) f(1,1,1). 
\label{ftof2l}
\end{equation}
Please note that in this example the recursion stopped due to condition~(\ref{g2lcond}),
so the result (in terms of $\alpha$) is exact;
in general one should stop when the desired order of $\alpha$ is reached
(for compactness we do not expand $g(\underline{n})$ into a series in $\alpha$ explicitly).
Also note that only even powers of $\alpha$ (integer powers of $d$) contribute 
to the final relation (\ref{ftof2l}), unlike the relations (\ref{ftog2l2}, \ref{gtog2l}).

The standard Laporta's approach, that is solving eqs.~(\ref{l2rr}, \ref{f2lcond}) for some (sufficiently large) set of 
values for $(\underline{n})$, gives
\begin{eqnarray}
0&=&R_{11}(1,1,1)-R_{21}(1,1,1) \nonumber\\ 
&=&1/2 (d - 2 \overline{\bf I}_1^- \overline{\bf I}_1^+  -\overline{\bf I}_2^- \overline{\bf I}_2^+  + \overline{\bf I}_2^+  +\dots)f(1,1,1)\nonumber\\
&=& (d/2 - 3/2) f(1,1,1) + 1/2 f(2,1,1), \nonumber
\end{eqnarray}
which confirms eq.~(\ref{ftof2l}).

Calculating the parameters $\lambda_a$ is the key point 
for obtaining recursive relations for coefficients of $1/d$ series.
To find $\lambda_a$, it is necessary to solve eqs.~(\ref{condla}), 
that is, a system of quadratic equations for $L(L+1)/2$ variables 
(for example, 28 variables for the 7-loop case).
However, the bilinear structure of eqs.~(\ref{condla}) allows 
us to significantly reduce this number.
First, using the linear equation (\ref{Trcondla}) we can exclude one of $\lambda_a$.
Moreover, we can obtain additional linear relations on $\lambda_a$
checking general combinations of eqs.~(\ref{condla}) with polynomials in $\lambda_a$
as coefficients. 
If such relations are found, with their help we can eliminate
some $\lambda_a$ from eqs.~(\ref{condla}) and
repeat the same steps for the resulting equations.

\begin{figure}[tbp]
\centering 
\begin{picture}(90,70) 

\put(0,60){$L=5$}
\put(30, 0){\line( 1, 0){30}} \put(43, 2){\small $c$}
\put(30, 0){\line( 0, 1){60}} \put(33,42){\small $e$} \put(33,14){\small $e$}
\put(30,60){\line( 1, 0){30}} \put(43,53){\small $c$}
\put(60,60){\line( 0,-1){60}} \put(53,42){\small $e$} \put(53,14){\small $e$}

\put(0,30){\line( 1, 1){30}} \put( 9,45){\small $b$}
\put(0,30){\line( 1,-1){30}} \put( 9,10){\small $b$}
\put(1,29){\line( 1, 0){20}} \put(10,33){\small $a$}
\put(1,31){\line( 1, 0){20}}

\put(90,30){\line(-1, 1){30}} \put(77,45){\small $b$}
\put(90,30){\line(-1,-1){30}} \put(77,10){\small $b$}
\put(89,29){\line(-1, 0){20}}
\put(89,31){\line(-1, 0){20}}

\put(30,30){\line( 1, 0){30}} \put(43,32){\small $c$}
\end{picture}
\hskip 2em
\begin{picture}(90,70)

\put(0,60){$L=6$}
\put(30, 0){\line( 1, 0){30}} \put(43, 2){\small $c$}
\put(30, 0){\line( 0, 1){60}} 
\put(33,10){\small $e$}\put(33,28){\small $e$}\put(33,45){\small $e$}
\put(30,60){\line( 1, 0){30}} \put(43,62){\small $c$}
\put(60,60){\line( 0,-1){60}}
\put(53,10){\small $e$}\put(53,28){\small $e$}\put(53,45){\small $e$}

\put(0,30){\line( 1, 1){30}} \put( 9,45){\small $b$}
\put(0,30){\line( 1,-1){30}} \put( 9,10){\small $b$}
\put(1,29){\line( 1, 0){20}} \put(10,33){\small $a$}
\put(1,31){\line( 1, 0){20}}

\put(90,30){\line(-1, 1){30}} \put(77,45){\small $b$}
\put(90,30){\line(-1,-1){30}} \put(77,10){\small $b$}
\put(89,29){\line(-1, 0){20}}
\put(89,31){\line(-1, 0){20}}

\put(30,20){\line( 1, 0){30}} \put(43,22){\small $c$}
\put(30,40){\line( 1, 0){30}} \put(43,42){\small $c$}
\end{picture}
\hskip 2em 
\begin{picture}(90,70) 

\put(0,60){$L=7$}
\put(30, 0){\line( 1, 0){30}} \put(43, 2){\small $c$}
\put(30, 0){\line( 0, 1){60}} 
\put(23, 9){\small $e$}\put(23,17){\small $e$}
\put(23,32){\small $e$}\put(23,47){\small $e$}
\put(30,60){\line( 1, 0){30}} \put(43,62){\small $c$}
\put(60,60){\line( 0,-1){60}}
\put(62, 9){\small $e$}\put(62,17){\small $e$}
\put(62,32){\small $e$}\put(62,47){\small $e$}

\put(0,30){\line( 1, 1){30}} \put( 9,45){\small $b$}
\put(0,30){\line( 1,-1){30}} \put( 9,10){\small $b$}
\put(1,29){\line( 1, 0){20}} \put(10,33){\small $a$}
\put(1,31){\line( 1, 0){20}}

\put(90,30){\line(-1, 1){30}} \put(77,45){\small $b$}
\put(90,30){\line(-1,-1){30}} \put(77,10){\small $b$}
\put(89,29){\line(-1, 0){20}}
\put(89,31){\line(-1, 0){20}}

\put(30,15){\line( 1, 0){30}} \put(43,17){\small $c$}
\put(30,30){\line( 1, 0){30}} \put(43,32){\small $c$}
\put(30,45){\line( 1, 0){30}} \put(43,47){\small $c$}
\end{picture}
\vskip 1em
\begin{picture}(120,70) 
\put(0,60){\small $L=6$}

\put(30, 0){\line( 1, 0){60}} 
\put(39, 2){\small $e$} \put(58, 2){\small $e$} \put(76, 2){\small $e$}
\put(30,60){\line( 1, 0){60}}
\put(39,62){\small $e$} \put(58,62){\small $e$} \put(76,62){\small $e$}
\put(30, 0){\line( 1, 1){26}}
\put(90,60){\line(-1,-1){26}}  
\put(90, 0){\line(-1, 1){26}}
\put(30,60){\line( 1,-1){26}} 
\put(33,47){\small $c$}\put(48,47){\small $c$}\put(60,47){\small $c$}\put(72,47){\small $c$}

\put(50, 0){\line( 1, 3){9}}
\put(70, 0){\line(-1, 3){9}}
\put(50,60){\line( 1,-3){9}} 
\put(70,60){\line(-1,-3){9}}

\put(15,30){\line( 1, 2){15}} \put(14,42){\small $b$}
\put(15,30){\line( 1,-2){15}} \put(14,14){\small $b$}
\put(16,29){\line( 1, 0){88}} \put(30,33){\small $a$}
\put(16,31){\line( 1, 0){88}}

\put(105,30){\line(-1, 2){15}} \put(101,42){\small $b$}
\put(105,30){\line(-1,-2){15}} \put(101,14){\small $b$}
\end{picture}
\hskip 1em
\begin{picture}(120,70) 
\put(0,60){\small $L=7$}

\put(30, 0){\line( 1, 0){60}}
\put(37, 2){\small $e$}\put(51, 2){\small $e$}\put(64, 2){\small $e$}\put(79, 2){\small $e$}
\put(30,60){\line( 1, 0){60}}
\put(37,62){\small $e$}\put(51,62){\small $e$}\put(64,62){\small $e$}\put(79,62){\small $e$}
\put(30, 0){\line( 1, 1){26}}
\put(90,60){\line(-1,-1){26}} 
\put(90, 0){\line(-1, 1){26}}
\put(30,60){\line( 1,-1){26}} 
\put(33,47){\small $c$}\put(44,47){\small $c$}\put(53,47){\small $c$}
\put(63,47){\small $c$}\put(72,47){\small $c$}

\put(45, 0){\line( 1, 2){13}}
\put(60, 0){\line( 0, 1){26}}
\put(75, 0){\line(-1, 2){13}}
\put(45,60){\line( 1,-2){13}}
\put(60,60){\line( 0,-1){26}}
\put(75,60){\line(-1,-2){13}}

\put(15,30){\line( 1, 2){15}} \put(14,42){\small $b$}
\put(15,30){\line( 1,-2){15}} \put(14,14){\small $b$}
\put(16,29){\line( 1, 0){88}} \put(30,33){\small $a$}
\put(16,31){\line( 1, 0){88}}

\put(105,30){\line(-1, 2){15}} \put(101,42){\small $b$}
\put(105,30){\line(-1,-2){15}} \put(101,14){\small $b$}
\end{picture}

\caption{\label{Fig:ALL}
$\overline{\lambda}_a$ (roots of eqs.~(\ref{condla}))
for some vacuum diagrams. Ordinary lines are massless, double line is massive with $\mu=1$;
\hbox{$a_L=L$}, \hbox{$b_L=-\frac{L-2}{2}e_L$}, 
\hbox{$c_L=\frac{L(L-2)}{2}e_L-L(L-1)$};\\
for planar diagrams (first row)
\hbox{$e_5= \frac{20}{9}$},  \hbox{$e_6= \frac{15}{8}$}, 
\hbox{$e_7=(\pm 28\,(61)^{1/2} + 1232)/625$};\\
for nonplanar (second row)
\hbox{$e_6=(\pm 3\,(41)^{1/2} + 57)/16$},
\hbox{$e_7=(\pm 168\,(26)^{1/2} + 2268)/625$}.
}
\end{figure}
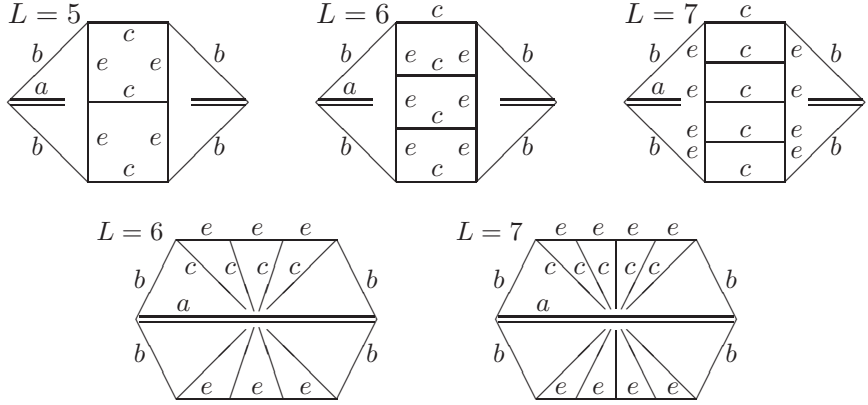

Figure \ref{Fig:ALL} presents ``hard'' $\overline{\lambda}_a$
(``soft'' $\tilde{\lambda}_a$ can be recovered using eqs.~(\ref{A2hatA}))
for leading sectors of some integral families up to the 7-loop order.
Polynomials of the first degree as coefficients were sufficient 
to reduce the original system of equations to an equation for one 
variable.\footnote{Symbolic manipulations were performed using REDUCE \cite{FreeREDUCE}).}

The results show that in the 5-, 6-loop planar examples
there is at least one variant of the substitutions
which transform IBP relations to a recursive form, in the 7-loop planar 
and in the 6-, 7-loop nonplanar cases there are at least two.
The $L$-dependence of the results was obtained empirically, but there are some
indications of its validity for an arbitrary order,
we hope to clarify this point in future publications.

Please note that eqs.~(16) from \cite{Baikov:2005nv} 
(eqs.~(\ref{condlas1}) in the notation of this article)
at the 7-loop level lead to polynomial equations 
of the 6th degree for 10 variables, which is
a much more difficult task.

\section{Conclusion}
\label{sec:Conc}
In this paper we present a method to construct
solutions of IBP relations for Feynman integrals as formal series
in the limit $1/d\rightarrow0$.
The idea is to use substitutions of the form (\ref{ftog},~\ref{ftog0})
which transform IBP relations into an explicitly recursive form.
From a technical point of view, it is necessary to calculate the constants $\lambda$
as solutions to a system of bilinear quadratic equations (\ref{condla}).
Although this is relatively simple (and therefore can be automated) in the case 
where one line is massive and the others are massless, more general 
kinematic configurations may be more difficult to study.
In particular, it may happen that the number of $1/d$ solutions constructed 
will be less than the number of master integrals and, 
therefore, not enough to calculate all $c_i(\underline{n})$ coefficients in eq.~(\ref {f2M}).
However, even in this case they provide some linear
relations between $c_i(\underline{n})$ that can be used (in combination with other
data) for their reconstruction or at least for verification.

Please note that although in this article we are dealing with recurrence relations for vacuum
integrals, the substitutions (\ref{ftog}, \ref{ftog0}) can also be
applied to the case with external momenta.
In this context, it is worth mentioning the work \cite{Mizera:2019vvs},
where the large $d$ limit was used to obtain the differential equations 
for master integrals.
We hope that studying possible transformations of the corresponding IBP relations
will help in the difficult task of evaluating ``multi-legs'' integrals;
we leave this for future research.

\providecommand{\href}[2]{#2}\begingroup\raggedright\endgroup


\begin{thebibliography}{99}

\bibitem{Chetyrkin:1981qh}
K.G.~Chetyrkin and F.V.~Tkachov, \emph{{Integration by Parts: The Algorithm to
  Calculate beta Functions in 4 Loops}},
  \href{https://doi.org/10.1016/0550-3213(81)90199-1}{\emph{Nucl. Phys. B}
  {\bfseries 192} (1981) 159}.

\bibitem{Tkachov:1981wb}
F.V.~Tkachov, \emph{{A Theorem on Analytical Calculability of Four Loop
  Renormalization Group Functions}},
  \href{https://doi.org/10.1016/0370-2693(81)90288-4}{\emph{Phys. Lett. B}
  {\bfseries 100} (1981) 65}.

\bibitem{Gorishnii:1989gt}
S.G.~Gorishnii, S.A.~Larin, L.R.~Surguladze and F.V.~Tkachov, \emph{{Mincer:
  Program for Multiloop Calculations in Quantum Field Theory for the
  Schoonschip System}},
  \href{https://doi.org/10.1016/0010-4655(89)90134-3}{\emph{Comput. Phys.
  Commun.} {\bfseries 55} (1989) 381}.

\bibitem{Anastasiou:2004vj}
C.~Anastasiou and A.~Lazopoulos, \emph{{Automatic integral reduction for higher
  order perturbative calculations}},
  \href{https://doi.org/10.1088/1126-6708/2004/07/046}{\emph{JHEP} {\bfseries
  07} (2004) 046} [\href{https://arxiv.org/abs/hep-ph/0404258}{{\ttfamily
  hep-ph/0404258}}].

\bibitem{Smirnov:2019qkx}
A.V.~Smirnov and F.S.~Chuharev, \emph{{FIRE6: Feynman Integral REduction with
  Modular Arithmetic}},
  \href{https://doi.org/10.1016/j.cpc.2019.106877}{\emph{Comput. Phys. Commun.}
  {\bfseries 247} (2020) 106877}
  [\href{https://arxiv.org/abs/1901.07808}{{\ttfamily 1901.07808}}].

\bibitem{Klappert:2020nbg}
J.~Klappert, F.~Lange, P.~Maierh\"ofer and J.~Usovitsch, \emph{{Integral
  reduction with Kira 2.0 and finite field methods}},
  \href{https://doi.org/10.1016/j.cpc.2021.108024}{\emph{Comput. Phys. Commun.}
  {\bfseries 266} (2021) 108024}
  [\href{https://arxiv.org/abs/2008.06494}{{\ttfamily 2008.06494}}].

\bibitem{Lee:2012cn}
R.N.~Lee, \emph{{Presenting LiteRed: a tool for the Loop InTEgrals REDuction}},
   \href{https://arxiv.org/abs/1212.2685}{{\ttfamily 1212.2685}}.

\bibitem{Studerus:2009ye}
C.~Studerus, \emph{{Reduze-Feynman Integral Reduction in C++}},
  \href{https://doi.org/10.1016/j.cpc.2010.03.012}{\emph{Comput. Phys. Commun.}
  {\bfseries 181} (2010) 1293}
  [\href{https://arxiv.org/abs/0912.2546}{{\ttfamily 0912.2546}}].

\bibitem{vonManteuffel:2012np}
A.~von Manteuffel and C.~Studerus, \emph{{Reduze 2 - Distributed Feynman
  Integral Reduction}},  \href{https://arxiv.org/abs/1201.4330}{{\ttfamily
  1201.4330}}.

\bibitem{Ruijl:2017cxj}
B.~Ruijl, T.~Ueda and J.A.M.~Vermaseren, \emph{{Forcer, a FORM program for the
  parametric reduction of four-loop massless propagator diagrams}},
  \href{https://doi.org/10.1016/j.cpc.2020.107198}{\emph{Comput. Phys. Commun.}
  {\bfseries 253} (2020) 107198}
  [\href{https://arxiv.org/abs/1704.06650}{{\ttfamily 1704.06650}}].

\bibitem{Georgoudis:2021onj}
A.~Georgoudis, V.~Gon\c{c}alves, E.~Panzer, R.~Pereira, A.V.~Smirnov and
  V.A.~Smirnov, \emph{{Glue-and-cut at five loops}},
  \href{https://doi.org/10.1007/JHEP09(2021)098}{\emph{JHEP} {\bfseries 09}
  (2021) 098} [\href{https://arxiv.org/abs/2104.08272}{{\ttfamily
  2104.08272}}].

\bibitem{Chakraborty:2022yan}
A.~Chakraborty, T.~Huber, R.N.~Lee, A.~von Manteuffel, R.M.~Schabinger,
  A.V.~Smirnov et~al., \emph{{Hbb vertex at four loops and hard matching
  coefficients in SCET for various currents}},
  \href{https://doi.org/10.1103/PhysRevD.106.074009}{\emph{Phys. Rev. D}
  {\bfseries 106} (2022) 074009}
  [\href{https://arxiv.org/abs/2204.02422}{{\ttfamily 2204.02422}}].

\bibitem{Lee:2023dtc}
R.N.~Lee, A.~von Manteuffel, R.M.~Schabinger, A.V.~Smirnov, V.A.~Smirnov and
  M.~Steinhauser, \emph{{Master integrals for four-loop massless form
  factors}}, \href{https://doi.org/10.1140/epjc/s10052-023-12179-2}{\emph{Eur.
  Phys. J. C} {\bfseries 83} (2023) 1041}
  [\href{https://arxiv.org/abs/2309.00054}{{\ttfamily 2309.00054}}].

\bibitem{Behring:2020cqi}
A.~Behring, F.~Buccioni, F.~Caola, M.~Delto, M.~Jaquier, K.~Melnikov et~al.,
  \emph{{Mixed QCD-electroweak corrections to $W$-boson production in hadron
  collisions}}, \href{https://doi.org/10.1103/PhysRevD.103.013008}{\emph{Phys.
  Rev. D} {\bfseries 103} (2021) 013008}
  [\href{https://arxiv.org/abs/2009.10386}{{\ttfamily 2009.10386}}].

\bibitem{Bern:2023ccb}
Z.~Bern, E.~Herrmann, R.~Roiban, M.S.~Ruf, A.V.~Smirnov, V.A.~Smirnov et~al.,
  \emph{{Conservative Binary Dynamics at Order $\alpha^5$ in Electrodynamics}},
  \href{https://doi.org/10.1103/PhysRevLett.132.251601}{\emph{Phys. Rev. Lett.}
  {\bfseries 132} (2024) 251601}
  [\href{https://arxiv.org/abs/2305.08981}{{\ttfamily 2305.08981}}].

\bibitem{Baikov:2005nv}
P.A.~Baikov, \emph{{A Practical criterion of irreducibility of multi-loop
  Feynman integrals}},
  \href{https://doi.org/10.1016/j.physletb.2006.01.052}{\emph{Phys. Lett. B}
  {\bfseries 634} (2006) 325}
  [\href{https://arxiv.org/abs/hep-ph/0507053}{{\ttfamily hep-ph/0507053}}].

\bibitem{Baikov:2007zza}
P.A.~Baikov, \emph{{Recurrence relations in the large space-time dimension
  limit}}, \href{https://doi.org/10.22323/1.048.0022}{\emph{PoS} {\bfseries
  RADCOR2007} (2007) 022}.

\bibitem{Baikov:2008jh}
P.A.~Baikov, K.G.~Chetyrkin and J.H.~Kuhn, \emph{{Order alpha**4(s) QCD
  Corrections to Z and tau Decays}},
  \href{https://doi.org/10.1103/PhysRevLett.101.012002}{\emph{Phys. Rev. Lett.}
  {\bfseries 101} (2008) 012002}
  [\href{https://arxiv.org/abs/0801.1821}{{\ttfamily 0801.1821}}].

\bibitem{Baikov:2009bg}
P.A.~Baikov, K.G.~Chetyrkin, A.V.~Smirnov, V.A.~Smirnov and M.~Steinhauser,
  \emph{{Quark and gluon form factors to three loops}},
  \href{https://doi.org/10.1103/PhysRevLett.102.212002}{\emph{Phys. Rev. Lett.}
  {\bfseries 102} (2009) 212002}
  [\href{https://arxiv.org/abs/0902.3519}{{\ttfamily 0902.3519}}].

\bibitem{Baikov:2010je}
P.A.~Baikov, K.G.~Chetyrkin and J.H.~Kuhn, \emph{{Adler Function, Bjorken Sum
  Rule, and the Crewther Relation to Order $\alpha^4_s$ in a General Gauge
  Theory}}, \href{https://doi.org/10.1103/PhysRevLett.104.132004}{\emph{Phys.
  Rev. Lett.} {\bfseries 104} (2010) 132004}
  [\href{https://arxiv.org/abs/1001.3606}{{\ttfamily 1001.3606}}].

\bibitem{Baikov:2012er}
P.A.~Baikov, K.G.~Chetyrkin, J.H.~Kuhn and J.~Rittinger, \emph{{Complete ${\cal
  O}(\alpha_s^4)$ QCD Corrections to Hadronic $Z$-Decays}},
  \href{https://doi.org/10.1103/PhysRevLett.108.222003}{\emph{Phys. Rev. Lett.}
  {\bfseries 108} (2012) 222003}
  [\href{https://arxiv.org/abs/1201.5804}{{\ttfamily 1201.5804}}].

\bibitem{Baikov:2012zm}
P.A.~Baikov, K.G.~Chetyrkin, J.H.~Kuhn and J.~Rittinger, \emph{{Vector
  Correlator in Massless QCD at Order $\mathcal{O}(\alpha^4_s)$ and the QED
  beta-function at Five Loop}},
  \href{https://doi.org/10.1007/JHEP07(2012)017}{\emph{JHEP} {\bfseries 07}
  (2012) 017} [\href{https://arxiv.org/abs/1206.1284}{{\ttfamily 1206.1284}}].

\bibitem{Baikov:2012zn}
P.A.~Baikov, K.G.~Chetyrkin, J.H.~Kuhn and J.~Rittinger, \emph{{Adler Function,
  Sum Rules and Crewther Relation of Order $\mathcal{O}(\alpha^4_s)$: the
  Singlet Case}},
  \href{https://doi.org/10.1016/j.physletb.2012.06.052}{\emph{Phys. Lett. B}
  {\bfseries 714} (2012) 62} [\href{https://arxiv.org/abs/1206.1288}{{\ttfamily
  1206.1288}}].

\bibitem{Baikov:2014qja}
P.A.~Baikov, K.G.~Chetyrkin and J.H.~K\"uhn, \emph{{Quark Mass and Field
  Anomalous Dimensions to ${\cal O}(\alpha_s^5)$}},
  \href{https://doi.org/10.1007/JHEP10(2014)076}{\emph{JHEP} {\bfseries 10}
  (2014) 076} [\href{https://arxiv.org/abs/1402.6611}{{\ttfamily 1402.6611}}].

\bibitem{Baikov:2016tgj}
P.A.~Baikov, K.G.~Chetyrkin and J.H.~K\"uhn, \emph{{Five-Loop Running of the
  QCD coupling constant}},
  \href{https://doi.org/10.1103/PhysRevLett.118.082002}{\emph{Phys. Rev. Lett.}
  {\bfseries 118} (2017) 082002}
  [\href{https://arxiv.org/abs/1606.08659}{{\ttfamily 1606.08659}}].

\bibitem{Baikov:2017ujl}
P.A.~Baikov, K.G.~Chetyrkin and J.H.~K\"uhn, \emph{{Five-loop fermion anomalous
  dimension for a general gauge group from four-loop massless propagators}},
  \href{https://doi.org/10.1007/JHEP04(2017)119}{\emph{JHEP} {\bfseries 04}
  (2017) 119} [\href{https://arxiv.org/abs/1702.01458}{{\ttfamily
  1702.01458}}].

\bibitem{Baikov:1996rk}
P.A.~Baikov, \emph{{Explicit solutions of the three loop vacuum integral
  recurrence relations}},
  \href{https://doi.org/10.1016/0370-2693(96)00835-0}{\emph{Phys. Lett. B}
  {\bfseries 385} (1996) 404}
  [\href{https://arxiv.org/abs/hep-ph/9603267}{{\ttfamily hep-ph/9603267}}].

\bibitem{Baikov:1996iu}
P.A.~Baikov, \emph{{Explicit solutions of the multiloop integral recurrence
  relations and its application}},
  \href{https://doi.org/10.1016/S0168-9002(97)00126-5}{\emph{Nucl. Instrum.
  Meth. A} {\bfseries 389} (1997) 347}
  [\href{https://arxiv.org/abs/hep-ph/9611449}{{\ttfamily hep-ph/9611449}}].

\bibitem{Laporta:2000dsw}
S.~Laporta, \emph{{High-precision calculation of multiloop Feynman integrals by
  difference equations}},
  \href{https://doi.org/10.1142/S0217751X00002159}{\emph{Int. J. Mod. Phys. A}
  {\bfseries 15} (2000) 5087}
  [\href{https://arxiv.org/abs/hep-ph/0102033}{{\ttfamily hep-ph/0102033}}].

\bibitem{FreeREDUCE}
{Anthony C. Hearn and Rainer Schöpf}, \emph{{REDUCE User’s Manual Free Version}}, 
\href{https://reduce-algebra.sourceforge.io/manual/manual.pdf}
{https://reduce-algebra.sourceforge.io/manual/manual.pdf (2024)}.

\bibitem{Mizera:2019vvs}
S.~Mizera and A.~Pokraka, \emph{{From Infinity to Four Dimensions: Higher
  Residue Pairings and Feynman Integrals}},
  \href{https://doi.org/10.1007/JHEP02(2020)159}{\emph{JHEP} {\bfseries 02}
  (2020) 159} [\href{https://arxiv.org/abs/1910.11852}{{\ttfamily
  1910.11852}}].

\end{thebibliography}
\end{document}